\documentclass[notitlepage,twocolumn,prd,showpacs,preprintnumbers,nofootinbib,superscriptaddress,floatfix,tightenlines]{revtex4-1}
\pdfoutput=1
\usepackage{mathrsfs}
\usepackage{amssymb}
\usepackage{amsmath}
\usepackage{graphicx}

\usepackage{hyperref}
\usepackage{amsfonts,amsbsy}
\usepackage{color}

\makeatletter
  \def\l@subsubsection#1#2{}%
\makeatother

\begin{document}

\title{Basin of attraction for turbulent thermalization and\\ the range of validity of classical-statistical simulations}

\author{J.~Berges}
\affiliation{Institut f\"{u}r Theoretische Physik, Universit\"{a}t Heidelberg, Philosophenweg 16, 69120 Heidelberg, Germany}
\affiliation{ ExtreMe Matter Institute (EMMI), GSI Helmholtzzentrum f\"ur Schwerionenforschung GmbH, 
Planckstra\ss e~1, 64291~Darmstadt, Germany}

\author{K.~Boguslavski}
\email{K.Boguslavski@thphys.uni-heidelberg.de}
\affiliation{Institut f\"{u}r Theoretische Physik, Universit\"{a}t Heidelberg, Philosophenweg 16, 69120 Heidelberg, Germany}

\author{S.~Schlichting}
\affiliation{Brookhaven National Laboratory, Physics Department, Bldg.\ 510A, Upton, NY 11973, USA}

\author{R.~Venugopalan}
\affiliation{Brookhaven National Laboratory, Physics Department, Bldg.\ 510A, Upton, NY 11973, USA}

\begin{abstract}
Different thermalization scenarios for systems with large fields have been proposed in the literature based on classical-statistical lattice simulations approximating the underlying quantum dynamics. We investigate the range of validity of these simulations for condensate driven as well as fluctuation dominated initial conditions for the example of a single component scalar field theory. We show that they lead to the same phenomenon of turbulent thermalization for the whole range of (weak) couplings where the classical-statistical approach is valid. In the turbulent regime we establish the existence of a dual cascade characterized by universal scaling exponents and scaling functions. This complements previous investigations where only the direct energy cascade has been studied for the single component theory. A proposed alternative thermalization scenario for stronger couplings is shown to be beyond the range of validity of classical-statistical simulations. 
\end{abstract}

\maketitle

\section{Introduction}

A quantum many-body system in thermal equilibrium is independent of its history in time and characterized by a few conserved quantities only. Therefore any thermalization process starting from a nonequilibrium initial state requires an effective loss of details about the initial conditions at sufficiently long times. It was pointed out previously that an effective {\it partial memory loss} of initial condition details can already be observed at earlier stages for the nonequilibrium unitary time evolution in quantum field theory~\cite{Berges:2004ce}. 

An extreme case occurs if at some early stage the nonequilibrium dynamics becomes {\it self-similar}. This amounts to an enormous reduction of the sensitivity to details of the underlying theory and initial conditions. The time evolution in this self-similar regime is described in terms of universal scaling exponents and scaling functions. While normalizations of the latter depend on model parameters such as couplings, masses and initial conditions, their functional form is universal. This has the important practical consequence that the nonequilibrium dynamics for whole ranges of different model parameters and initial conditions can be grouped into universality classes. 
The self-similar time-evolution within a given universality class of different models can then be mapped onto each other by simple rescalings. The universal properties are described in terms of nonthermal fixed points~\cite{Berges:2008wm} in renormalization group theory~\cite{Berges:2008sr}. While the notion of thermal fixed points describing different universality classes is well established for systems in thermal equilibrium, the corresponding classification for nonequilibrium scaling phenomena is a rapidly progressing research topic.  

Universal behavior far from equilibrium has been predicted for systems ranging from early-universe inflaton dynamics to table-top experiments with cold atoms. In these examples, attractor solutions with self-similar scaling behavior are associated with the phenomenon of Kolmogorov wave turbulence~\cite{Micha:2002ey,Micha:2004bv} and strong turbulence with fluid-like behavior~\cite{Berges:2008wm,Berges:2010ez,Nowak:2010tm,Nowak:2011sk} which leads to Bose condensation far from equilibrium~\cite{Berges:2012us,Nowak:2012gd,Gasenzer:2013era}. 
A turbulent thermalization mechanism has also been predicted for non-Abelian gauge theories using classical-statistical simulations in a fixed box, where different groups obtain consistent results~\cite{Berges:2008mr,Schlichting:2012es,Kurkela:2012hp}.

Recently, the existence of a nonthermal fixed point in longitudinally expanding non-Abelian plasmas was demonstrated using classical-statistical lattice simulations in the weak coupling limit~\cite{Berges:2013eia,Berges:2013fga}. It was argued that heavy-ion collision experiments at sufficiently high energies can provide the relevant initial conditions in the basin of attraction of this nonthermal fixed point. At sufficiently high energies, the colliding nuclei are described in the Color Glass Condensate framework~\cite{Gelis:2010nm}. The dynamics of the nonequilibrium Glasma~\cite{Lappi:2006fp} created in such a collision is that of gluon fields with typical momentum Q and a weak gauge coupling $\alpha_s(Q)$. Since the characteristic occupancies $\sim 1/\alpha_s(Q)$ are large, the gauge fields are strongly correlated even for small gauge coupling and the Glasma exhibits classical-statistical dynamics. 

The early-time behavior is then described by plasma instabilities, which have been intensively analyzed over the past years~\cite{Instab1,Instab2,Instab3,Instab4}. Classical-statistical simulations show that gluons become highly occupied $\sim 1/\alpha_s(Q)$ after a proper time $\sim Q^{-1} \log^2 (\alpha_s^{-1})$ for expanding systems~\cite{Romatschke:2006nk,Fukushima:2011nq,Berges:2012cj}. Subsequently, as argued in~\cite{Berges:2013eia,Berges:2013fga}, one expects a transition to the universal weak coupling attractor. 

However, in Ref.~\cite{Gelis:2013rba} it has recently been argued that by exceeding a rather small value of the gauge coupling $\alpha = g^2/4\pi$ with $g \sim 0.5$, the dynamics changes dramatically compared to previous weak-coupling estimates, using small initial Gaussian distributed fluctuations superimposed to the large classical macroscopic field $\sim {\mathcal O}(1/g)$ as derived in Ref.~\cite{Epelbaum:2013waa}. Moreover, similar findings of pressure isotropization for condensate driven initial conditions have been reported in nonequilibrium classical-statistical scalar field theory with expansion~\cite{Dusling:2012ig}. 

Even without expansion, a different thermalization scenario in the presence of large initial condensates was discussed using classical-statistical simulations once the self-coupling $\lambda$ reaches a certain strength~\cite{Epelbaum:2011pc}. In contrast, in the weak-coupling limit, the same initial conditions are known to lead to high occupancies $\sim 1/\lambda$ after a time $t \sim Q^{-1} \log (\lambda^{-1})$ where the characteristic scale $Q$ is determined by the energy density of the system~\cite{Kofman:1994rk,Berges:2002cz}. 
The subsequent evolution exhibits the phenomenon of turbulent thermalization~\cite{Micha:2002ey} reminiscent of the classical-statistical non-Abelian gauge theory results in the weak coupling limit~\cite{Berges:2008mr,Schlichting:2012es,Kurkela:2012hp}. The important advantage for scalars is that powerful resummation techniques exist for the description of the entire evolution in quantum field theory. It has been demonstrated that classical-statistical descriptions have a well understood range of validity for which they accurately describe the dynamics of the underlying quantum theory~\cite{Aarts:2001yn,Moore:2001zf,Arrizabalaga:2004iw,Berges:2008wm}. 

In this work, we revisit the question of nonequilibrium dynamics from condensate driven initial conditions for the single component theory employed in Ref.~\cite{Epelbaum:2011pc}. We first investigate the dynamics in the weak coupling limit. In this limit, we observe turbulent thermalization, where we point out that 
\begin{itemize}
\item[1.] the inverse particle cascade, which has so far only been established for $O(N)$ symmetric theories with $N \ge 2$ \cite{Berges:2008wm,Berges:2010ez,Gasenzer:2011by}, also occurs for the single component theory.
\end{itemize}
This finding directly complements the investigations of Refs.~\cite{Micha:2002ey,Micha:2004bv}, where the direct energy cascade has been studied for the single component theory while the phenomenon of a dual cascade was not considered. We then explore the basin of attraction for the nonthermal fixed point by comparing two generic classes of initial conditions:  
\begin{itemize}
\item[2.] Condensate driven initial conditions with large field and small (vacuum) fluctuations,
\item[3.] fluctuation dominated initial conditions with small (zero) field and large fluctuations,
\end{itemize}
and confirm~\cite{Micha:2002ey,Micha:2004bv} that they lead to the same universal behavior beyond a time scale $\sim Q^{-1} \log(\lambda^{-1})$ for the whole range of couplings where the classical-statistical approach is valid. Thus the observed differences in thermalization scenarios that have been proposed are not caused by differences in the initial conditions. We finally investigate  
\begin{itemize}
\item[4.] the range of validity of classical-statistical simulations
\end{itemize}
and explain why the results of Ref.~\cite{Epelbaum:2011pc} do not lie in this range.

\section{Scalar model and initial conditions}

There is a restricted class of problems where the dynamics of bosonic quantum fields can be accurately mapped onto a classical-statistical problem. The most intuitive criteria for this can be formulated in situations where a kinetic description in terms of quasi-particle excitations is applicable. The system exhibits classical dynamics whenever the typical occupation numbers per mode are much larger than unity,
\begin{equation}
f(t,p) \, \gg \, 1 \, . 
\label{eq:classicality}
\end{equation}
If occupation numbers fall below unity, quantum processes will dominate the dynamics. This is clearly seen in a Boltzmann transport framework where classical scattering processes are sub-leading to quantum ones for occupation numbers smaller than unity~\cite{Mueller:2002gd,Jeon:2004dh}.

It is also possible to formulate more general criteria, which do not rely on a quasi-particle picture, in the Schwinger-Keldysh formalism of nonequilibrium quantum field theory~\cite{Aarts:2001yn,Berges:2004yj,Berges:2007ym,Jeon:2013zga}. This `classicality condition' is met whenever anti-commutator expectation values for typical bosonic field modes are much larger than the corresponding commutators \cite{Aarts:2001yn,Berges:2007ym}. 
Stated differently, this concerns the large field or large occupancy limit, which is relevant for important phenomena such as nonequilibrium instabilities or wave turbulence encountered in our study. 
The classicality condition has been verified in detail by comparing quantum to classical-statistical results in the context of scalar quantum field dynamics \cite{Aarts:2001yn,Berges:2007ym,Arrizabalaga:2004iw,Berges:2008wm} and coupled to fermions~\cite{Berges:2010zv,Berges:2013oba}.

We emphasize that the condition for a system to exhibit classical dynamics is in general time dependent. In particular, the approach to complete thermal equilibrium is not accessible within the classical-statistical framework. As a consequence of the Rayleigh-Jeans divergence, the classical thermal state is only well defined for an ultraviolet cutoff $\Lambda$, which may be implemented by a lattice regularization. In its range of validity, observables computed from classical-statistical simulations are insensitive to the Rayleigh-Jeans divergence. Thermal equilibrium is a genuine quantum state which cannot be reached within classical-statistical field theory. Nevertheless, the classical-statistical regime may extend over large times such that the quantitative properties of the nonequilibrium quantum evolution are accurately described within the classical-statistical approach.

The time scale $t_{\rm quant}$ for entering the quantum regime is not a universal quantity and depends in general on the properties of the initial state as well as the dynamics in the classical regime. For the considered cases of large initial fields or large initial fluctuations at weak coupling, this time scale is parametrically given by~$t_{\rm quant}\sim Q^{-1} \lambda^{-5/4}$ in our case~\cite{Micha:2004bv}. The range of validity of classical-statistical techniques in time is thus naturally confined to weak couplings. In general the use of classical-statistical methods at large couplings requires great care since genuine quantum effects may dominate the dynamics already at rather early times. This will be investigated in detail below after we present the physical weak coupling results.

We study the dynamics of a massless real scalar field theory with quartic interaction. The classical action is given by
\begin{eqnarray}
S\left[ \varphi \right] = \int d^4x \left( \frac{1}{2} \partial_{\mu}\varphi \partial^{\mu}\varphi-\frac{\lambda}{24} 
\varphi^4 \right) \;,
\label{mat:class-action-static}
\end{eqnarray}
summing over $ \mu = 0,1,2,3$ in Minkowski space-time. The classical equation of motion $\delta S/\delta \varphi = 0$ for the single-component field reads
\begin{eqnarray}
  \square \varphi(x) + \frac{\lambda}{6}\varphi^3(x) = 0\;.
 \label{mat:class-EOM-static}
\end{eqnarray}

Following Ref.~\cite{Epelbaum:2011pc}, we consider Gaussian initial conditions. They are fully determined by the initial one- and two-point correlation functions of the fields $\varphi(x)$ and $\pi(x) = \partial \varphi(x)/\partial x^0$ at time $t = x^0 = 0$. For a spatially homogeneous system the most general initial Gaussian conditions can be characterized by
\begin{eqnarray}
 \phi(t=0) = \langle \varphi(0,{\bf x}) \rangle  &,& \dot{\phi}(0)  =  \langle \pi(0,{\bf x}) \rangle
 \label{eq:phiinitial}\\
 F(t=t'=0,{\bf x}-{\bf y}) &=& \langle \varphi(0,{\bf x}) \varphi(0,{\bf y}) \rangle - \phi^2(0) 
 \\
 K(t=t'=0,{\bf x}-{\bf y}) &=& \langle \pi(0,{\bf x}) \pi(0,{\bf y}) \rangle - \dot{\phi}^2(0) 
\end{eqnarray} 
as well as\footnote{The antisymmetric combination $\sim \langle \varphi({\bf x}) \pi({\bf y}) - \pi({\bf x}) \varphi({\bf y}) \rangle$ is fixed by the corresponding field commutation relation in the quantum theory (or Poisson bracket in the classical theory)~\cite{Berges:2004yj}.} 
\begin{eqnarray} 
 \frac{1}{2}\langle \varphi(0,{\bf x}) \pi(0,{\bf y}) + \pi(0,{\bf x}) \varphi(0,{\bf y}) \rangle - \phi(0) {\dot\phi}(0) \,. 
 \label{eq:phipiini}
\end{eqnarray} 
The values for these initial correlation functions are taken to coincide with those from the corresponding quantum initial value problem. In the classical-statistical theory the correlation functions are obtained by performing a phase-space average over the initial configurations for a given observable
\begin{eqnarray}
\langle O_{\textrm{cl}}(\varphi,\pi) \rangle = \int \mathcal{D}\varphi(0)\mathcal{D}\pi(0)W(\varphi(0),\pi(0))O_{\textrm{cl}}(\varphi,\pi) \, 
 \label{mat:class-stat-average}
\end{eqnarray}
with respect to the distribution function of initial fields $W(\varphi(0),\pi(0))$. The latter is chosen to give the prescribed correlation functions (\ref{eq:phiinitial}--\ref{eq:phipiini}). Accordingly, we sample the initial field configurations until convergence to the prescribed initial correlations is achieved. Each initial configuration is separately evolved in time and the time evolution of correlation functions is obtained from the ensemble average. 

Using the corresponding definitions also for $t > 0$, we may extract the time evolution of occupation numbers $f(t,p)$ in spatial Fourier space with $p = |{\bf p}|$ from~\cite{Berges:2004yj}
\begin{eqnarray}
 f(t,p) + \frac{1}{2} = \sqrt{\tilde{F}(t,p)\tilde{K}(t,p)} \, ,
 \label{eq:distribution}
\end{eqnarray}
as well as the dispersion
\begin{eqnarray}
 \omega(t,p) = \sqrt{\frac{\tilde{K}(t,p)}{\tilde{F}(t,p)}} \, ,
\end{eqnarray}
where $\tilde{F}(t,p) = \int d^3 x~e^{-i {\bf p} {\bf x}} F(t,x)$ denotes the Fourier transform. With these definitions, the above initial conditions specify the initial field amplitude $\phi(t=0)$ and distribution function $f(t=0,p)$. The initial frequency reads $\omega(t=0,p)= \sqrt{p^2 + m^2_{\rm eff}}$, where the effective mass term is given by 
\begin{eqnarray}
 m_{\textrm{eff}}^2 = \frac{\lambda}{2}  \left( \phi^2(t=0) + \int_p^\Lambda F(t=0,p) \right) + \delta m^2_\Lambda \, .
 \label{eq:effective-mass}
\end{eqnarray}
In this setup the initial correlator $(\ref{eq:phipiini})$ vanishes and we will also employ $\dot{\phi}(t=0)=0$ in the following. 

The counter-term $\delta m^2_\Lambda$ in Eq.~(\ref{eq:effective-mass}) can be chosen to cancel the leading quadratic $\Lambda$-dependence of the three-dimensional momentum integral over the initial $F(t=0,p)$. The same counter term is then used to cancel the associated divergence in the equation of motion (\ref{mat:class-EOM-static}). If not stated otherwise, we will perform such renormalized simulations. In addition, we will also present results for $\delta m^{2}_{\Lambda}=0$ as employed in Ref.~\cite{Epelbaum:2011pc}.

In the classical equation of motion (\ref{mat:class-EOM-static}) all model dependence on the coupling constant $\lambda$ could be scaled out by the reparametrization $\varphi \rightarrow \varphi/\sqrt{\lambda}$. The corresponding information is then entirely encoded in the initial conditions. The coupling drops out everywhere except for the `quantum half' in the initial conditions, which would become `$\lambda/2$'. Therefore, the coupling `controls' the size of quantum corrections in the initial conditions~\cite{preheat1}. This also reflects the fact that the classical-statistical theory will accurately describe the quantum physics as long as the coupling is sufficiently small. We do not rescale the field in the following for the purpose of our presentation. 

The conserved total energy density $\epsilon$ is used to calculate the characteristic scale 
\begin{eqnarray}
 Q = \sqrt[4]{\lambda ~\epsilon}\;.
 \label{mat:Q-effective}
\end{eqnarray}
For the initial conditions, which we consider in the following, this scale is independent of the coupling constant.

The first set of initial conditions is characterized by a large macroscopic field 
\begin{eqnarray}
\label{mat:condensate-IC}
&& \phi(t=0) = \frac{\sigma_0}{\sqrt{\lambda}} \quad , \quad f(t=0,p) = 0 \;, \\
&& \qquad \quad \quad (\textrm{Condensate IC}) \nonumber
\end{eqnarray}
with $\sigma_0$ of order one. For this initial condition the mode occupancies are zero such that all modes are  initialized with the vacuum `quantum half' up to the lattice ultraviolet cutoff as in Ref.~\cite{Epelbaum:2011pc}. 

The second class of initial conditions is taken to be fluctuation dominated for comparison, with an overoccupied distribution up to some initial momentum $Q_0$,
\begin{eqnarray}
\label{mat:box-IC}
&& \phi(t=0)=0\;, \quad f(p) = \frac{n_0}{\lambda}\, \Theta\!\left( Q_0 - p \right)\;, \\
&& \qquad \qquad \quad (\textrm{Fluctuation IC})  \nonumber
\end{eqnarray}
and the initial amplitude of the particle distribution $n_0$ with $Q \sim \sqrt[4]{n_0}~Q_0$. For both types of initial conditions one finds that the energy density scales with the coupling constant as 
$\epsilon \sim 1/\lambda$.

\section{Turbulent thermalization}

The condensate driven initial conditions (\ref{mat:condensate-IC}) lead to the well known phenomenon of parametric resonance. In the context of inflationary cosmology, the corresponding preheating dynamics~\cite{Kofman:1994rk} and the process of turbulent thermalization has been studied in great detail both using classical-statistical simulations~\cite{preheat1,preheat2,Micha:2002ey,Micha:2004bv,Berges:2008wm,Berges:2010ez,Berges:2012us} as well as directly in quantum field theory using resummation techniques based on the two-particle irreducible (2PI) effective action~\cite{Berges:2002cz,Arrizabalaga:2004iw,Berges:2008wm}. 
In the cosmological context, the preheating dynamics has also been investigated in the presence of fermion matter~\cite{ferm1,ferm2} and their consequences deduced for fermion production and the phenomenon of turbulent thermalization~\cite{Berges:2010zv,Berges:2013oba}.  
 
Initially, the energy density is stored in the large coherent field $\phi(t)$ for small coupling $\lambda$. 
Since $\phi(t)$ is rapidly oscillating, we will also consider the behavior of the envelope of the maximum field amplitude $\phi_0(t)$ as a function of time. For the subsequent evolution one can identify three characteristic ranges in time, which are parametrically given as follows.\\

1. {\it Instability} for $0 \ll t \ll Q^{-1} \log(\lambda^{-1})$: In this parametric resonance regime, the field $\phi(t)$ is a periodic oscillating function with frequency characterized by the initial rescaled field amplitude $\sigma_0$~\cite{Boyanovsky:1994me,Greene:1997fu}. This initial amplitude also determines the initial resonance band in momenta for which an exponential growth of $f(t,p)$ can be observed~\cite{Greene:1997fu}. The rapid growth of fluctuations leads to strong nonlinearities which broaden the resonance band and produce enhanced growth rates~\cite{Berges:2002cz}. As a consequence, a wide range of growing modes lead to a prethermalization of the equation of state at the end of this early stage~\cite{Berges:2004ce,Podolsky:2005bw,Dusling:2010rm} while the distribution function itself is still far from equilibrium.\\ 

2. {\it Turbulence} for $Q^{-1} \log(\lambda^{-1}) \ll t \ll Q^{-1} \lambda^{-5/4}$: In this regime, the field amplitude approaches a power-law behavior and the distribution function becomes self-similar~\cite{Micha:2004bv}. Elastic scattering processes dominate and a dual cascade forms in distinct momentum ranges: A direct energy cascade towards the ultraviolet modes~\cite{Micha:2004bv} and an inverse particle cascade towards the infrared develop~\cite{Berges:2008wm}. 
These cascades are separated in momentum space by the characteristic momentum of the dominant resonance peak in the distribution function. The direct energy cascade drives the thermalization process by energy transport towards higher momenta. The inverse particle cascade leads to the phenomenon of Bose condensation in this far from equilibrium regime~\cite{Berges:2012us}.\\

3. {\it Thermalization} for $t \gg Q^{-1} \lambda^{-5/4}$: In this regime elastic and inelastic processes lead to a Bose-Einstein distribution. This regime is beyond the range of validity of classical-statistical simulations. At sufficiently late times the classical evolution will always end up showing a classical thermal distribution with a temperature parameter $T_\Lambda$ depending on the Rayleigh-Jeans cutoff $\Lambda$. The inability of the classical approach to describe this late-time thermalization stage has been studied in detail in the literature~\cite{Aarts:2001yn,Berges:2004yj}.\\ 

The characteristic time scales are given as weak coupling parametric estimates for our purposes. In addition, some of the prefactors are also available that can be significantly different from one for the case of the single component scalar field theory. For instance, the time scale for the end of the instability regime is more accurately given by $t \simeq (2\gamma_0)^{-1} \log(\lambda^{-1})$, where $\gamma_0 \simeq 0.033 Q$~\cite{Greene:1997fu} is the largest growth rate of the primary instabilities. 

\begin{figure}[tp]
\centering
 \includegraphics[width=0.5\textwidth]{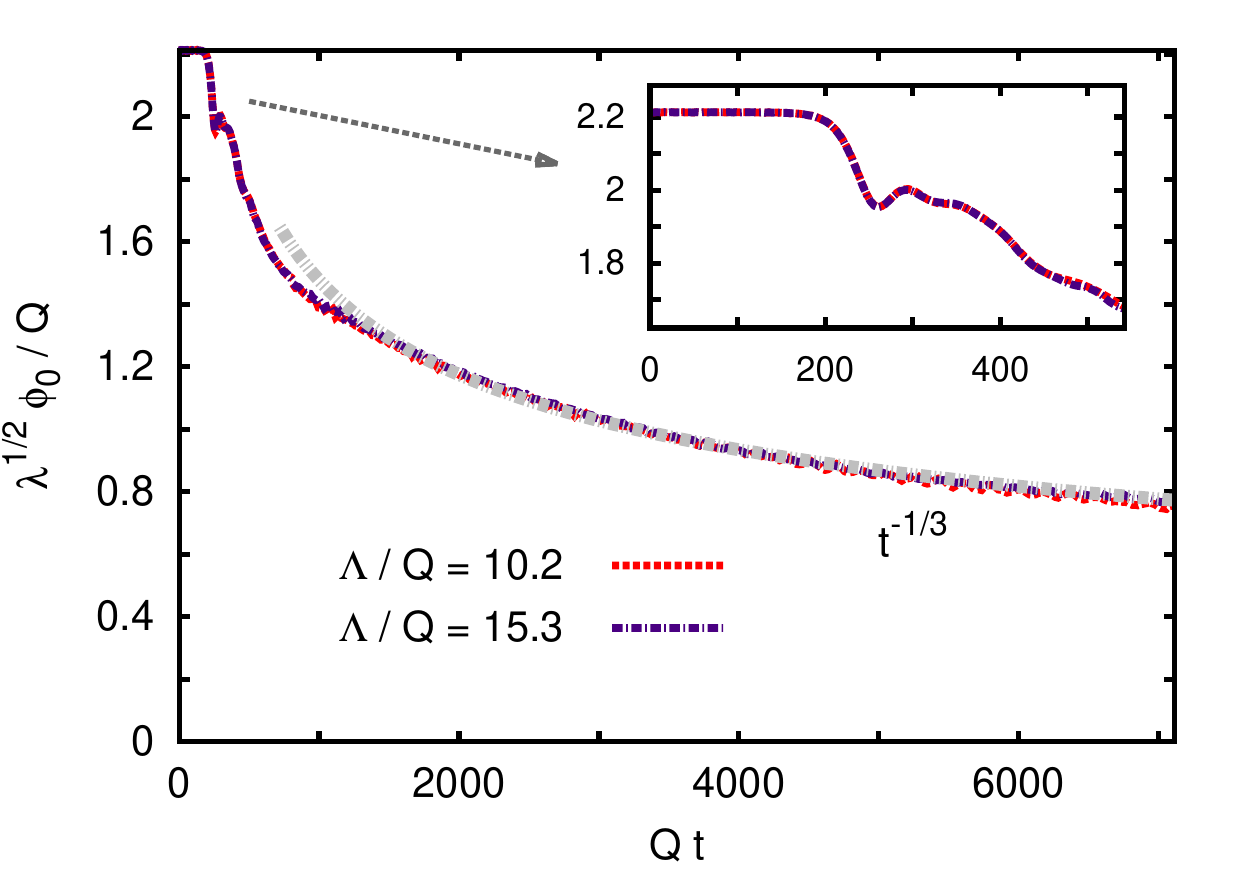}
 \caption{The rescaled field amplitude $\sqrt{\lambda}~\phi_0(t)$ for $\lambda = 10^{-4}$. We employ two different cutoffs in order to demonstrate the insensitivity of the results on the lattice regularization. Also shown is the predicted power law $\sim t^{-1/3}$ (gray dashed line), where one observes very good agreement with data at later times. In the inset, the same data is shown to better resolve early times.}
\label{fig:1e-4-condensate}
\end{figure}

The following numerical results are obtained using standard lattice discretization techniques~\cite{Berges:2004yj}. In Fig.~\ref{fig:1e-4-condensate} we show the time-dependent rescaled field amplitude $\sqrt{\lambda}~\phi_0(t)$ for a weak coupling $\lambda = 10^{-4}$ and $\sigma_0 = 2.21 Q$. Two different data sets are shown corresponding to two different lattice momentum cutoffs $\Lambda/Q = 10.2$ and $\Lambda/Q = 15.3$.\footnote{These results are obtained on $768^3$ lattices. The volumes are different for each cutoff but the lattice is large enough such that the evolution of the system is seen to be independent of the volume.} The comparison confirms that the weak coupling results are insensitive to the Rayleigh-Jeans cutoff in classical-statistical field theory, which is a crucial ingredient for its ability to describe the corresponding quantum field dynamics~\cite{Aarts:2001yn}.  

The oscillating behavior with constant maximum amplitude at early times is visible from the inset of Fig.~\ref{fig:1e-4-condensate}. Subsequently, for $t \sim Q^{-1} \log(\lambda^{-1})$ the corrections from classical-statistical fluctuations change this behavior dramatically and trigger a transient rapid field decay. This happens when the size of fluctuations has grown such that their contribution to the energy density becomes comparable to that from the macroscopic field. At this stage the evolution becomes strongly non-linear which even leads to a temporary field growth~\cite{Berges:2002cz}. The non-linear dynamics finally leads to a power-law decay of the field amplitude 
\begin{eqnarray}
 \phi_0(t) \sim Q~(Qt)^{-\delta}
\end{eqnarray}
with the predicted exponent $\delta = 1/3$~\cite{Micha:2004bv} given by the gray dashed line in Fig.~\ref{fig:1e-4-condensate}. 

As shown in Ref.~\cite{Micha:2004bv}, during the turbulent stage between $Q^{-1} \log(\lambda^{-1}) \ll t \ll Q^{-1} \lambda^{-5/4}$ the distribution function approaches a self-similar behavior,
\begin{eqnarray}
 f(t,p) = (Qt)^{\alpha} f_S((Qt)^{\beta} p)\;, 
 \label{mat:self-similar-static}
\end{eqnarray}
with the universal scaling exponents $\alpha$, $\beta$ and the stationary scaling function $f_S$. The dynamical scaling exponents describe the evolution of typical occupation numbers and momenta. For hard modes, which dominate the energy density, $\beta= -1/5$ determines the evolution of characteristic momenta and $\alpha=-4/5$ of their occupancy. The latter exponent determines the parametric time $t_{\rm quant}\sim Q^{-1} \lambda^{-5/4}$ at which the typical occupancies become order one and the classicality condition (\ref{eq:classicality}) is no longer fulfilled. 

Wave turbulence~\cite{bib:Zakharov-Wave-Turbulence} describes the transport of conserved quantities such as energy density. Accordingly, in momentum space power-law cascades form. This can be either a `direct cascade', for transport towards higher momenta, or an `inverse cascade' into the infrared. 
In the inertial range of momenta, where the distribution function is described as a momentum power-law, one can write
\begin{eqnarray}
 f(p) \sim \left(\frac{Q}{p}\right)^{\kappa}\;
\label{eq:fpower} 
\end{eqnarray}
with a universal scaling exponent $\kappa$. The proportionality factor depends in general on time for isolated systems as in our case, i.e.~without applied sources or sinks that could lead to stationary cascades.

If there is more than one conserved quantity, the system may accommodate this by forming distinct cascades in different momentum regimes. It has been shown that for characteristic momenta $p \lesssim Q$ the dynamics is dominated by elastic scatterings~\cite{Berges:2008wm}. This leads to an additional conserved quantity that emerges during the nonequilibrium time evolution even though total particle number is not conserved in the relativistic scalar field theory. As a consequence, apart from a direct energy cascade towards the ultraviolet, an inverse particle cascade towards the infrared can be observed. For the single component theory, this analysis was previously only done for the direct cascade~\cite{Micha:2002ey,Micha:2004bv}; the presence of the inverse cascade was not established.

\begin{figure}[tp]
\centering
 \includegraphics[width=0.5\textwidth]{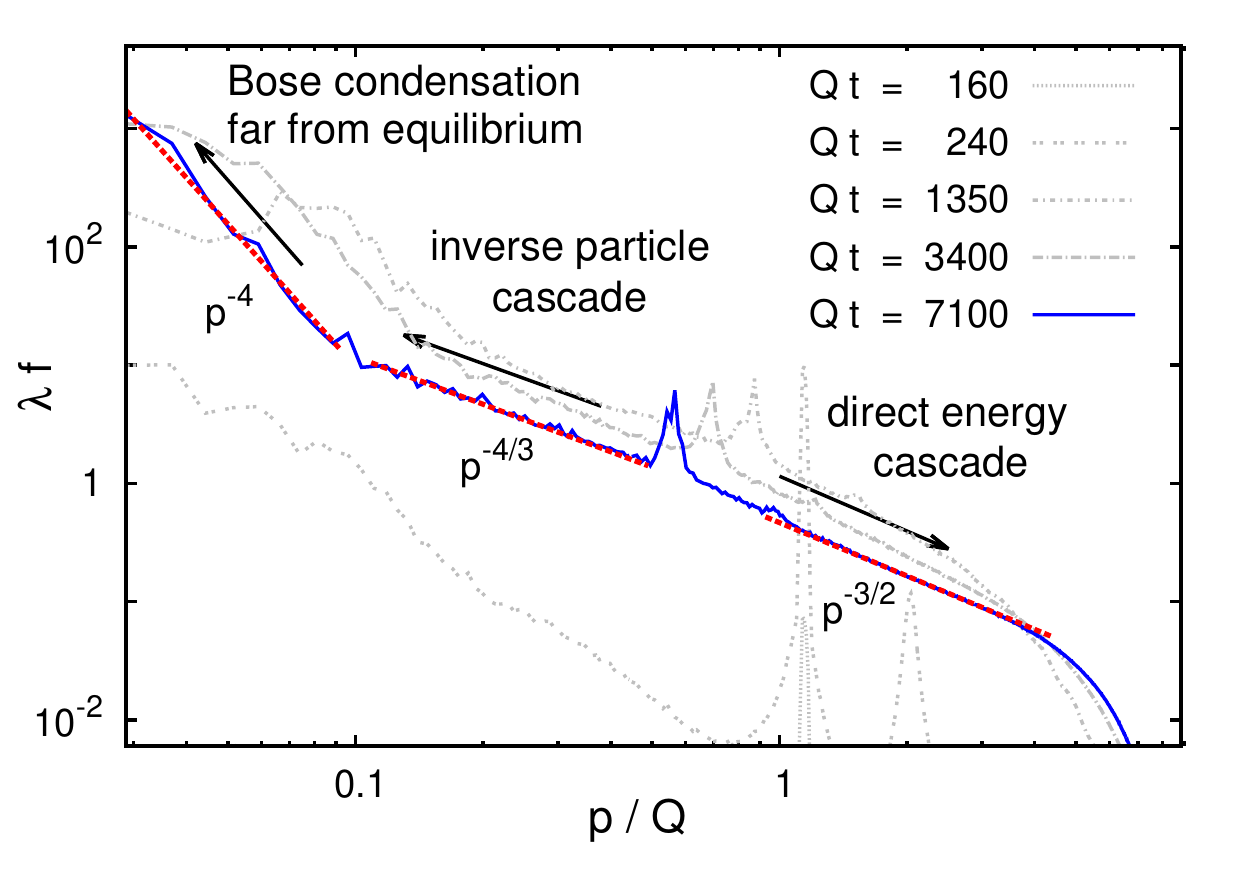}
 \caption{The rescaled distribution function $\lambda f$ at different times for $\lambda = 10^{-4}$. Power-laws with exponents $3/2$, $4/3$ and $4$ are also shown.}
\label{fig:1e-4}
\end{figure}

In Fig.~\ref{fig:1e-4} we show our results for the evolution of the rescaled distribution function $\lambda f(t,p)$ for different fixed times. At times $Qt=160$ and $240$ one observes the resonance behavior of the instability stage. The resonance peak occurs first around $p_*/Q \approx 1.1$ and its location is proportional to the amplitude of the condensate at early times, $p_* = 0.52 \sigma_0$~\cite{Greene:1997fu}.
Since the macroscopic field amplitude decreases with time, the peak is shifted to softer momenta~\cite{Micha:2004bv}. 

At later times, the peak serves as a source for energy and particles, leading to momentum-scale invariant energy and particle fluxes towards the ultraviolet and the infrared, respectively.
Accordingly, different momentum power-laws emerge to the left and to the right of the still visible dominant resonance peak in Fig.~\ref{fig:1e-4}. On the double logarithmic plot the power-laws are well described by straight lines with different slopes, corresponding to different values for the exponent of (\ref{eq:fpower}) in distinct momentum ranges. The spectrum at $Qt = 7100$ is compared to three power laws with the exponents 
\begin{eqnarray}
 \kappa_S = 4\;, \quad \kappa_M = \frac{4}{3}\;, \quad \kappa_H = \frac{3}{2}\;. 
\end{eqnarray}
The hard scale exponent $\kappa_H=3/2$ describes the energy cascade towards higher momenta and results from effective $2 \leftrightarrow (1 + \text{ soft})$ processes involving the condensate~\cite{Micha:2002ey,Micha:2004bv}. The inverse particle cascade towards lower momentum modes shows two distinct momentum regimes depending on the size of the occupation numbers $f(p)$ in each regime. For $\lambda f(p) \lesssim \mathcal{O}(1)$ the weak wave turbulence exponent $\kappa_M= 4/3$ describes the transport of particles. In the nonperturbative regime of ultrasoft momenta where $\lambda f(p) \gtrsim \mathcal{O}(1)$ the strong turbulence exponent $\kappa_S$ governs the dynamics~\cite{Berges:2008wm}. 

We emphasize that the entire inverse particle cascade, which is described by $\kappa_M$ and $\kappa_S$, can be understood in terms of elastic processes only. The enhanced scaling exponent $\kappa_S > \kappa_M$ is a consequence of an emergent effective scattering matrix element for $2 \leftrightarrow 2$ processes, which can be described in terms of a momentum dependent effective coupling $\lambda_{\rm eff}(p) \sim p^8$~\cite{Berges:2008wm,Berges:2010ez,Berges:2012us}. This analytical prediction is based on a systematic large-$N$ expansion to next to leading order of the 2PI effective action for the $N$-component scalar quantum field theory, which gives $\kappa_S = d + 1$ in $d$ space dimensions~\cite{Berges:2001fi}. Since we have $N = 1$, and in view of some reported deviations for $N=2$~\cite{Gasenzer:2011by}, it is remarkable that we find the predicted exponent with the observed accuracy.

In order to display the whole inverse particle cascade with both the weak and the strong turbulence regimes, we used rather large lattices up to $768^3$. Moreover, to cover the entire range of momenta shown in Fig.~2, we actually combined the data from two simulations with different lattice cutoffs $\Lambda/Q = 10.2$ and $\Lambda/Q = 15.3$. In the overlapping momentum regions the two simulations agree to very good accuracy. We emphasize that this is only done for presentational purposes of the physical weak coupling results. In particular, all results presented in the upcoming sections come from simulations including the entire range of displayed momenta.

\section{Basin of attraction for the nonthermal fixed point}

A crucial property of the turbulent regime is its strict independence of model parameters such as the value of the coupling constant $\lambda$ or the initial conditions such as the initial field value $\sigma_0$. This is a very powerful consequence of universality, which finds its manifestation in the self-similar behavior (\ref{mat:self-similar-static}). The latter represents an enormous reduction of the possible dependence of the dynamics on variations in time and momenta, since it states that $(Q t)^{-\alpha} f(t,p)$ only depends on the product $(Q t)^\beta p$ instead of separately depending on time and momenta. 
In general, renormalization group theory tells us that the fixed point distribution $f_S$ appearing in (\ref{mat:self-similar-static}) will depend on all 'relevant' parameters of the system. For instance, if the initial field value $\sigma_0$ would represent such a relevant parameter then $f_S$ would in addition depend on the product $(Q t)^{\zeta} \sigma_0$ with some new exponent $\zeta$. 
Plotting $(Q t)^{-\alpha} f(t,p)$ only as a function of $(Q t)^\beta p/Q$ would then fail to describe the data. Therefore (\ref{mat:self-similar-static}) represents a very strong statement about the loss of information about the parameters of the underlying system already at this transient stage of the nonequilibrium time evolution. 

Since self-similarity has been extensively discussed for our theory already in the literature~\cite{Micha:2004bv}, we only address here those aspects of universality that are relevant for clarifying the conflicting statements mentioned in the introduction. In particular, we want to confirm that both the class of condensate driven initial conditions given by (\ref{mat:condensate-IC}) as well as fluctuation dominated initial conditions (\ref{mat:box-IC}) belong to the basin of attraction for this nonthermal fixed point. 

\begin{figure}[tp]
\centering
 \includegraphics[width=0.5\textwidth]{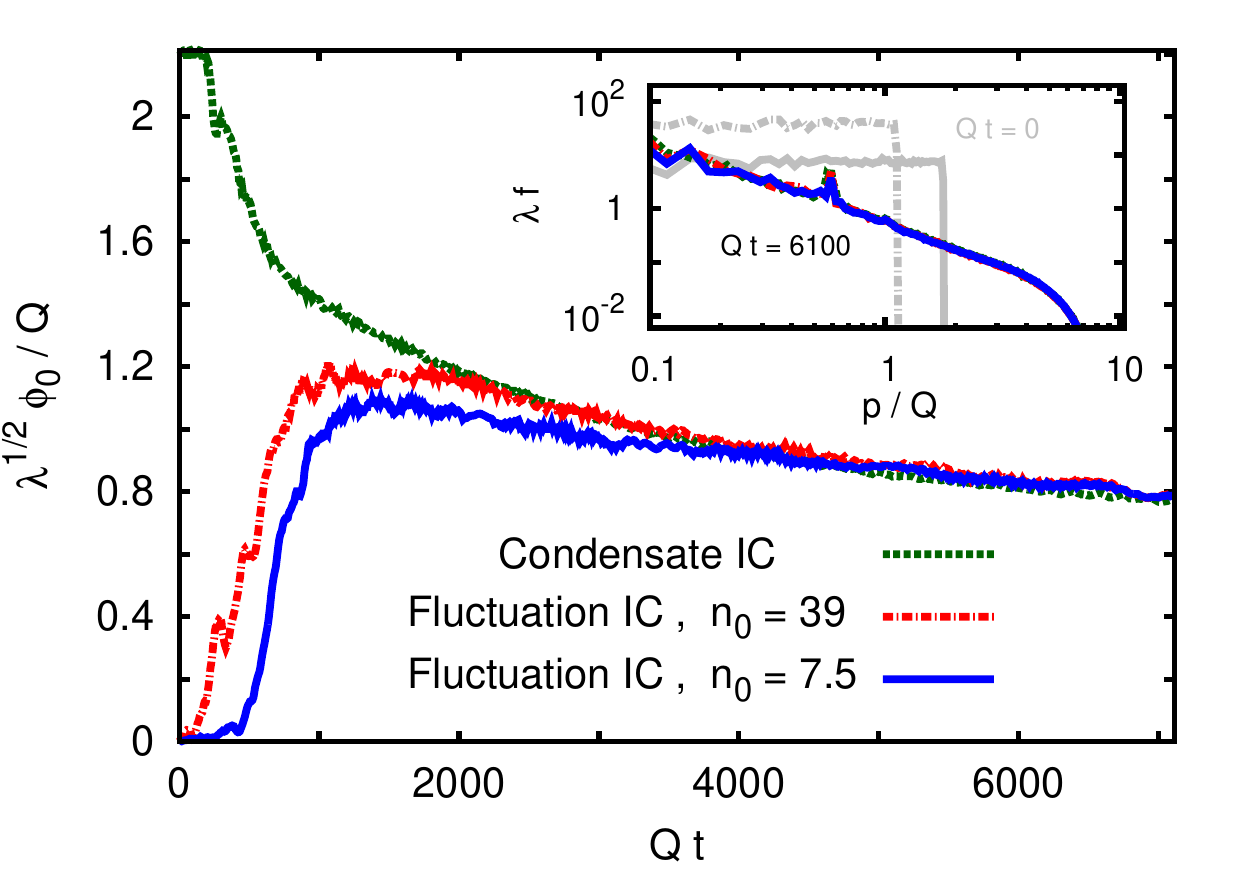}
 \caption{The rescaled field amplitude as a function of time for three different initial conditions at fixed coupling $\lambda = 10^{-4}$. The inset shows the corresponding occupation number distributions $\lambda f(t,p)$ at $Qt = 6100$. For the two fluctuation dominated initial conditions the inset shows in addition the initial spectra (gray). One observes that the evolution becomes independent of the initial conditions for the condensate driven as well as for the fluctuation dominated initializations.}
\label{fig:condensate-IC-L10-4}
\end{figure}

So far we have used the initial conditions characterized by a large field and small (vacuum) fluctuations.
The second class of initial conditions we now consider are described in terms of highly occupied modes
with initial amplitude $n_0 / \lambda$ up to the momentum $Q_0$. We consider $n_0 = 39$  as well as $n_0 = 7.5$.\footnote{The simulations for the fluctuation dominated initial conditions were performed on $256^3$ lattices, while for the condensate driven initial condition we employ a $768^3$ lattice but we checked that smaller lattices lead to the same observations. We use the lattice cutoff $\Lambda / Q = 15.3$, where we have verified that the results are insensitive to the cutoff.} The characteristic scale $Q$ given by (\ref{mat:Q-effective}) is taken to be always the same, thus fixing $Q_0$ for given $n_0$. We employ a weak coupling $\lambda = 10^{-4}$.

Using both types of initial conditions, the evolution of the rescaled field amplitude is shown in Fig.~\ref{fig:condensate-IC-L10-4}. One observes that for each of the fluctuation dominated initial conditions a condensate builds up\footnote{Any homogeneous field domain can only form as rapidly as allowed by causality.}, which is a consequence of the inverse particle cascade as described above~\cite{Berges:2012us}. Vice versa, the condensate driven initial conditions lead to fluctuations by parametric resonance. 
At late times the curves for all of the different initial conditions fall on top of each other to very good accuracy. This is further illustrated by showing the corresponding distribution functions in the inset of the figure. While the single particle spectra are very different at initial time $(Q t = 0)$, at late times the curves for all three distributions fall on top of each other. The fact that the evolution becomes independent of the details of the initial conditions already at this transient stage is a striking example of the phenomenon of universality far from equilibrium.

\section{Limitations of classical-statistical simulations}

We now come back to the condensate driven initial conditions, as employed in Ref.~\cite{Epelbaum:2011pc}, and explore the range of validity of classical-statistical simulations that provide a reliable description of the underlying quantum dynamics. Classical-statistical descriptions are restricted to the weak coupling regime as explained above. Therefore, it is important to verify up to which size of the coupling these methods can be applied.  
This is a time-dependent question since for sufficiently late times all classical-statistical descriptions break down once typical occupancies become order one -- as is the case for thermal equilibrium\footnote{An important exception concerns dynamic critical phenomena. Since the universal real-time properties at the critical point are controlled by the infrared dynamics, they can be accurately computed within the classical-statistical framework~\cite{Berges:2009jz}.} -- and genuine quantum processes dominate.\footnote{Here we do not consider the frequently employed possibility to initialize the 'quantum 1/2' only in the instability band~\cite{Arrizabalaga:2004iw}.} 

For weak couplings, the time at which the instability regime ends depends logarithmically on the inverse coupling constant as explained above. The subsequent evolution becomes universal and thus independent of the coupling. The system exits the turbulent regime and enters the quantum one around the parametric time $t_{\rm quant}\sim Q^{-1} \lambda^{-5/4}$. Increasing the coupling therefore means that the range of validity in time shrinks. Equivalently, for some fixed time one should observe deviations from universal behavior as the coupling is increased. The onset of any coupling dependence also signals the breakdown of the classical-statistical method. 

\begin{figure}[tp]
\centering
 \includegraphics[width=0.5\textwidth]{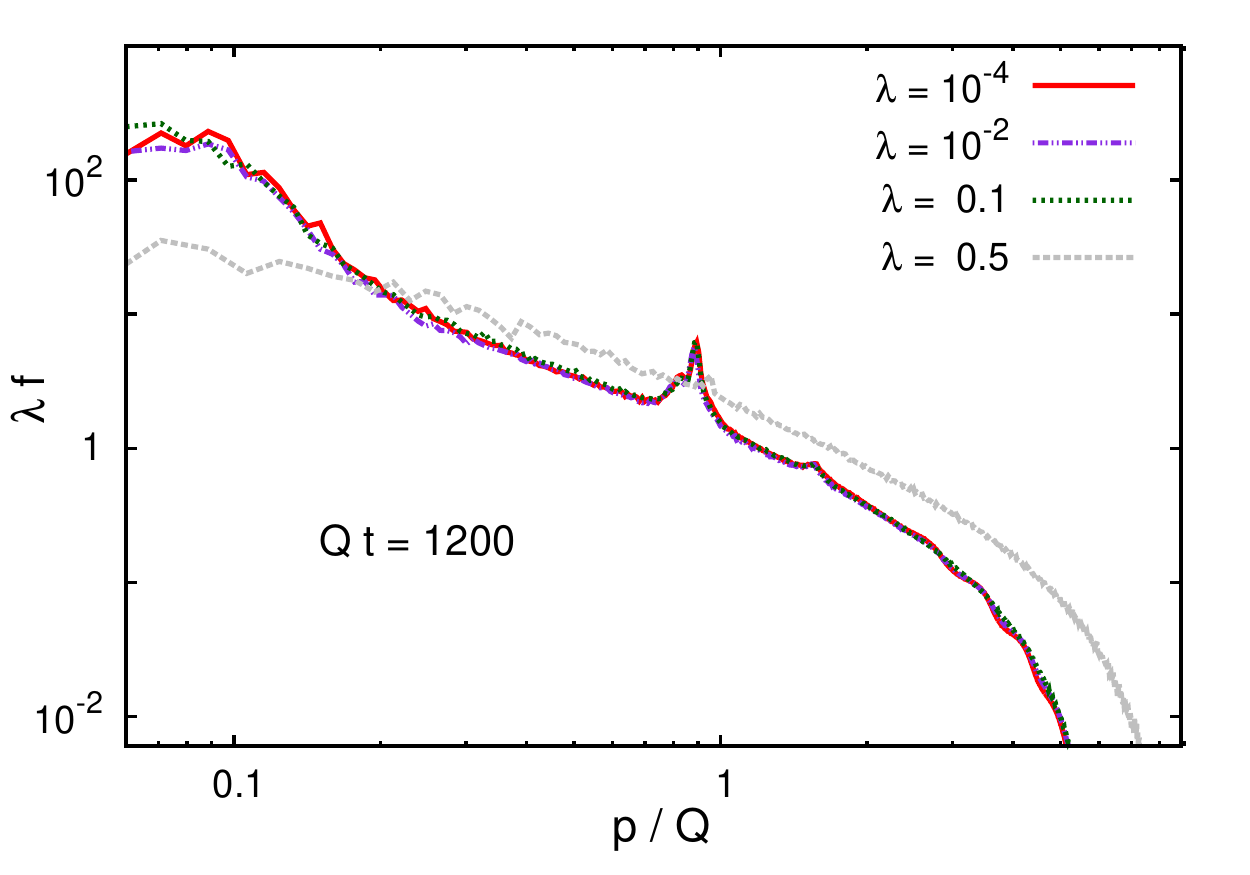}
 \caption{The rescaled distribution function $\lambda f(t,p)$ at time $Qt=1200$ for different values of the coupling $\lambda$. For $\lambda \lesssim 0.1$ the curves agree to very good accuracy by virtue of universality. Sizeable deviations can only be observed for larger coupling $\lambda=0.5$.}
\label{fig:VaryLWeak}
\end{figure}

This is illustrated in Fig.~\ref{fig:VaryLWeak} where we show the rescaled distribution function $\lambda f$ for different values of the coupling constant $\lambda$ at fixed time $Qt = 1200$. One observes that for weak couplings $\lambda \lesssim 0.1$ the spectra lie on top of each other to very good accuracy.\footnote{For a vanishing coupling constant ($\lambda = 0$) instabilities do not occur and the macroscopic field follows a classical evolution.}
In contrast, for $\lambda=0.5$ deviations are clearly visible and also the shape becomes qualitatively different, not showing the infrared enhancement or power law behavior. The presented results for $\lambda \leq 0.1$ are insensitive to the employed Rayleigh-Jeans cutoff. The $\lambda = 0.5$ curve is obtained for $\Lambda / Q = 15.3$.

\begin{figure}[tp]
\centering
 \includegraphics[width=0.5\textwidth]{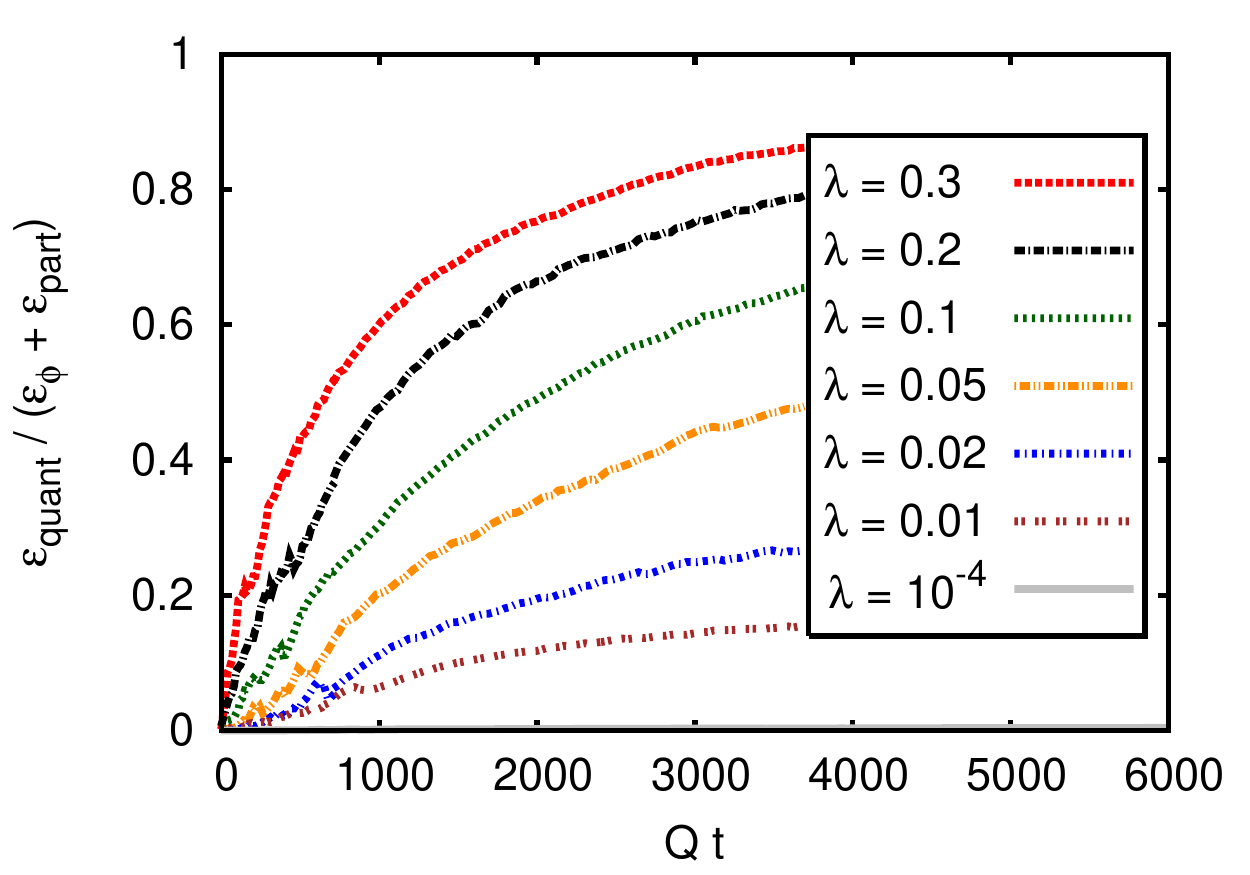}
 \caption{The quantum part of the energy density $\epsilon_{\textrm{quant}}$ as defined in (\ref{eq:epsquant}) divided by the sum of the condensate and fluctuation part given in the text as a function of time for different values of the coupling $\lambda$ for the cutoff $\Lambda/Q = 15.3$.}
\label{fig:EQu-over-ETotal}
\end{figure}

It is instructive to analyze the dynamics in this weak-coupling range in more detail. The classicality condition (\ref{eq:classicality}) implies that modes with low occupancies of $f(p) \lesssim 1$ should not play a dominant role for the dynamics. Their contribution to the energy density may be estimated by 
\begin{eqnarray}
 \epsilon_{\textrm{quant}}(t) = \int \frac{d^3 p}{(2 \pi)^3}~\omega(t,p)~f(t,p) ~\Theta \left( 1-f(t,p) \right).
 \label{eq:epsquant}
\end{eqnarray}
This `quantum' part should be compared to the total energy density contained in quasi-particle fluctuations $\epsilon_{\textrm{part}}(t) = \int_{\bf p} \omega(t,p) f(t,p)$ and in the macroscopic field, which we approximate by $\epsilon_\phi(t) = \lambda \phi_0^4(t)/24$. In the range of applicability of classical-statistical descriptions one expects that the ratio $\epsilon_{\textrm{quant}}/(\epsilon_\phi + \epsilon_{\textrm{part}}) \ll 1$. 
Of course, such a simple separation into fluctuation and field parts is only applicable if there is a good quasi-particle description for weak couplings and such an analysis should always be taken with great care. For instance, we will see below that for $\lambda = 1$ very strong cutoff dependence occurs and the occupation number distribution becomes negative such that this analysis is inapplicable. However, we find that for the range of couplings $\lambda \lesssim 0.2$ no strong cutoff dependence occurs and the distribution function fulfills $f(p) \ge 0$ in the classical-statistical regime. 

Fig.~\ref{fig:EQu-over-ETotal} shows this quantum part of the energy density divided by the sum of the condensate and fluctuation part as a function of time for different values of the coupling $\lambda$. While for $\lambda = 10^{-4}$ this ratio is practically zero for all displayed times, one observes that for $\lambda = 0.1$ it becomes already about one-third for $Q t = 1200$. From this time on, for $\lambda = 0.2$ the quantum part dominates the sum $\epsilon_\phi + \epsilon_{\textrm{part}}$ and a description in terms of classical-statistical simulations becomes questionable. These findings are in very good agreement with the above observation about emerging coupling dependencies of the results beyond the range of validity of the classical-statistical approach.

\section{Rayleigh-Jeans cutoff dependence at coupling $\lambda = 1$}

\begin{figure}[t]
\centering
 \includegraphics[width=0.5\textwidth]{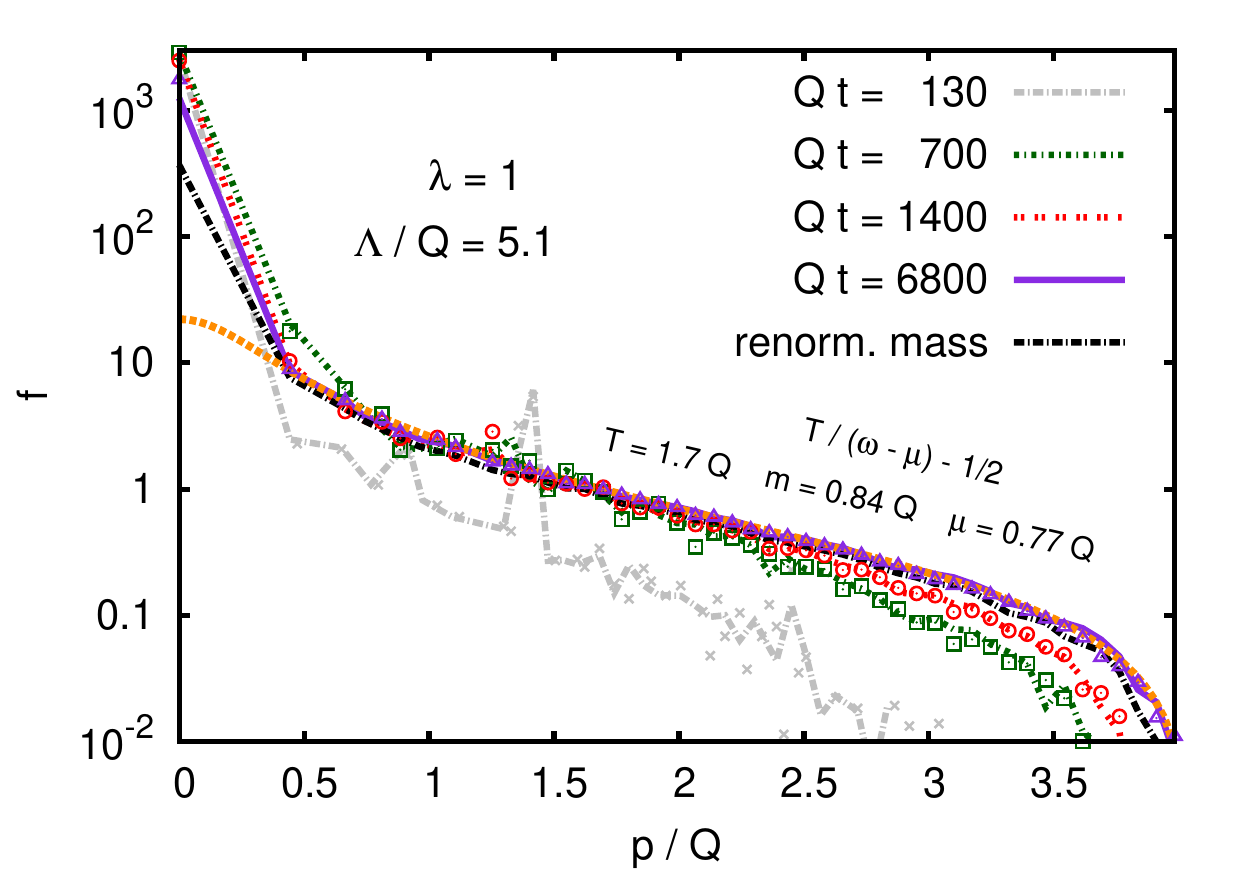}
 \caption{The occupation number distribution $f(t,p)$ for $\lambda=1$ at different times for the same parameters as in Ref.~\cite{Epelbaum:2011pc}. Our results (lines) agree to very good accuracy with the previous study (points). The black dashed curve gives the corresponding result with mass renormalization to show that no significant differences can be observed at $Q t = 6800$. The orange dashed curve is a fit to the spectrum as employed in Ref.~\cite{Epelbaum:2011pc} to extract temperature $T$, mass $m$ and chemical potential $\mu$ parameters.}
\label{fig:spectra-L1-A1-N20}
\end{figure}
 
\begin{figure}[t]
\centering
 \includegraphics[width=0.5\textwidth]{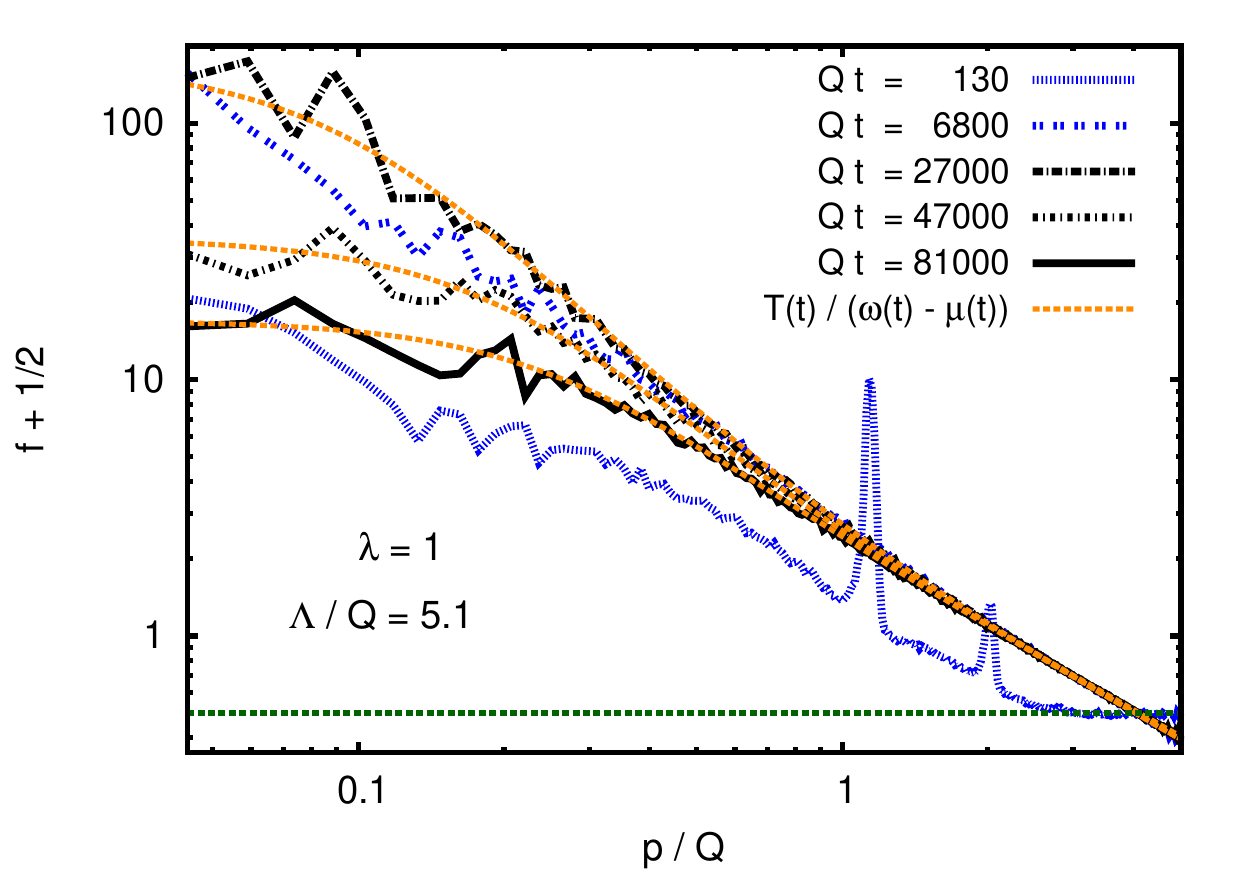}
 \caption{$f(t,p) + 1/2$ for $\lambda=1$ at different times with the same lattice cutoff as used in Fig.~\ref{fig:spectra-L1-A1-N20}. Blue curves denote growing occupancies at early times, black curves show the spectra at later times when occupation numbers are decreasing in time. The dashed orange lines are classical thermal functions, fitted to the black curves. The dashed green line denotes $f = 0$.}
\label{fig:spectra-L1-A1}
\end{figure}

We have seen above that quantum corrections become more important at later times. Therefore, one might hope that for short enough times one could extend the range of validity of classical-statistical simulations to  stronger couplings. However, since the coupling controls the relative size of the occupation number per mode as compared to the `quantum half' and the latter are cut off by the Rayleigh-Jeans regulator $\Lambda$, there is the danger that the sizeable cutoff dependence leads to a breakdown of classical-statistical simulations already at relatively early times in this case. 

In Ref.~\cite{Epelbaum:2011pc} it is argued that at short enough times and strong enough couplings one does not enter the weak coupling attractor for turbulent thermalization and an alternative thermalization scenario can be observed within the framework of classical-statistical simulations. The authors find that for $\lambda =1$ the classical-statistical thermalization dynamics is very different than what has been previously derived for weaker couplings. In particular, it is claimed that no turbulent cascades form and Bose condensation occurs as a consequence of the formation of a transient chemical potential. 

In the following we reconsider the calculations of Ref.~\cite{Epelbaum:2011pc}. First we also use their employed $20^3$ lattices and cutoff $\Lambda/Q = 5.1$ for the condensate driven initial conditions (\ref{mat:condensate-IC}) in the absence of mass renormalization ($\delta m^2_{\Lambda} = 0$). In Fig.~\ref{fig:spectra-L1-A1-N20} our results are plotted by lines along with the data from the referenced study given by points of the same colors. Both results agree to very good accuracy. We also give the corresponding curve with mass renormalization (black dashed curve) to show that no significant differences can be observed. 
The orange dashed curve is a fit to the spectrum as employed in Ref.~\cite{Epelbaum:2011pc}. One finds that $f(p)+1/2$ can be described by a classical Rayleigh-Jeans distribution $T/(\omega(p)-\mu)$ with temperature parameter $T$, frequency $\omega(p) = \sqrt{p^2 + m^2}$, (squared) mass parameter $m^2 = \frac{\lambda}{2} \left\langle \varphi^2 \right\rangle + \delta m^{2}_{\Lambda}$ and chemical potential $\mu$. 

We then extend these studies to larger lattices up to the size $512^3$ in order to also resolve the infrared physics. In particular, this will enable us to vary the Rayleigh-Jeans cutoff while being insensitive to finite volume effects. Our results with the same lattice cutoff but larger volume and mass renormalization are summarized in Fig.~\ref{fig:spectra-L1-A1}. The spectra start with the occupied quantum $1/2$. The blue curves denote earlier times when the distribution function grows with time. For $Q t = 130$, one finds the resonance structure while at $Q t = 6800$ hard momenta can already be fitted by a classical thermal distribution, as indicated in Fig.~\ref{fig:spectra-L1-A1-N20}.

The entire spectrum can be described by a thermal function starting from the time $Q t = 27000$. Going to even later times, we find that the classical thermal fitting function changes its parameters with time since the effective mass decreases. Therefore, the temperature and the chemical potential of the fitting function become time dependent, where $T$ grows while $\mu$ decreases. The difference between $\mu$ and $m$ grows and the chemical potential decreases faster than the effective mass. The values of the thermal fits denoted by dashed orange lines are $T/Q = 1.81$, $m/Q = 0.44$ and $\mu/Q = 0.43$ at $Q t = 27000$ and $T/Q = 1.92$, $m/Q = 0.41$ and $\mu/Q = 0.29$ at the time $Q t = 81000$.

\begin{figure}[t]
\centering
 \includegraphics[width=0.5\textwidth]{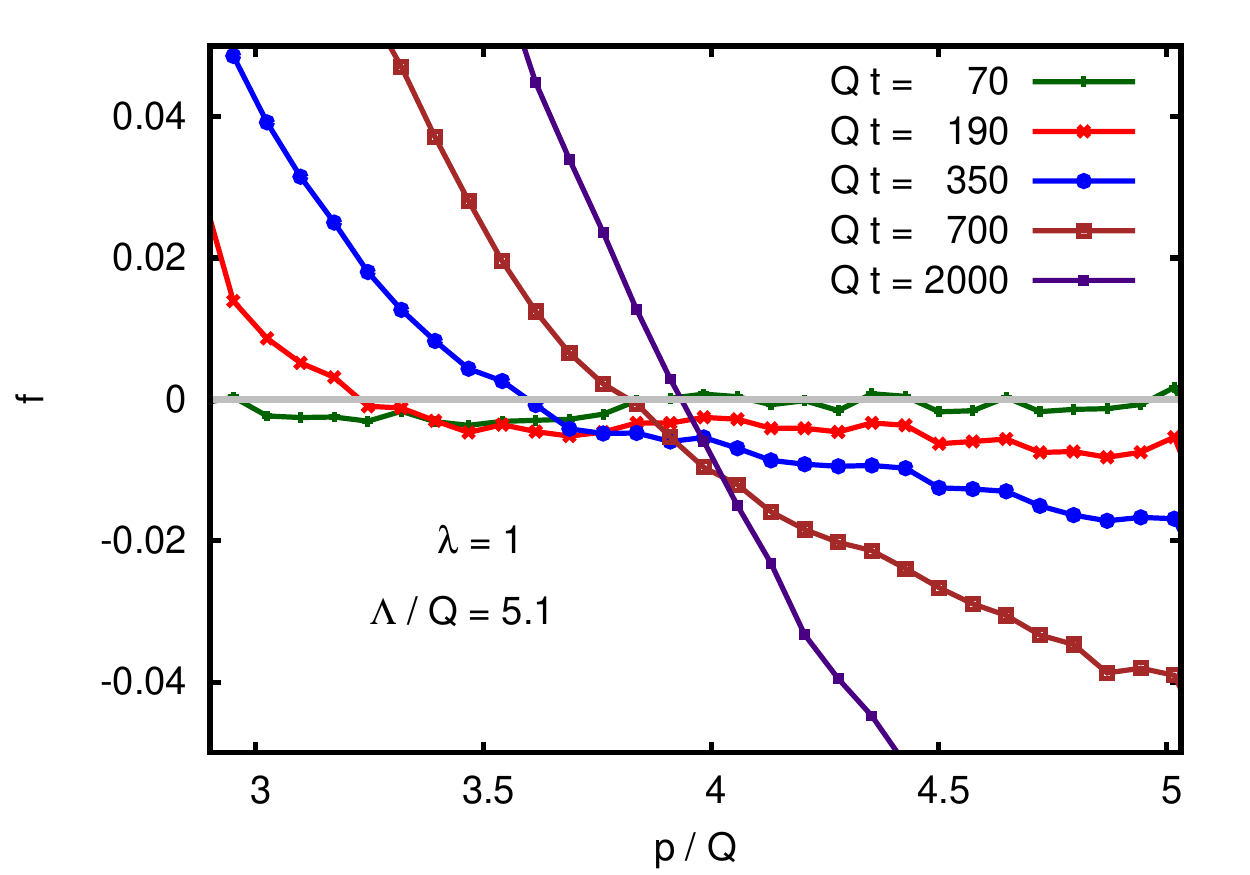}
 \caption{The tail of the spectrum $f$ at different times for lattice cutoff $\Lambda/Q = 5.1$ and coupling $\lambda = 1$. The gray line indicates $f=0$.}
\label{fig:negative-tail-L1}
\end{figure}

With classical equilibration in the total occupation number, the ultraviolet tail of the distribution function $f(p)$ becomes negative. Stated differently, the vacuum `quantum half' decays~\cite{Moore:2001zf}. This failure of the classical-statistical approximation to describe the otherwise stable vacuum of the quantum theory is illustrated in Fig.~\ref{fig:negative-tail-L1}, where we zoom into the hard momentum region. While for $\lambda=1$ this unphysical decay of the `quantum half' can already be observed at rather early times $Qt\sim100$, we emphasize that this does not happen in the weak coupling regime and clearly indicates the breakdown of the classical-statistical approach for large couplings.

\begin{figure}[t]
\centering
 \includegraphics[width=0.5\textwidth]{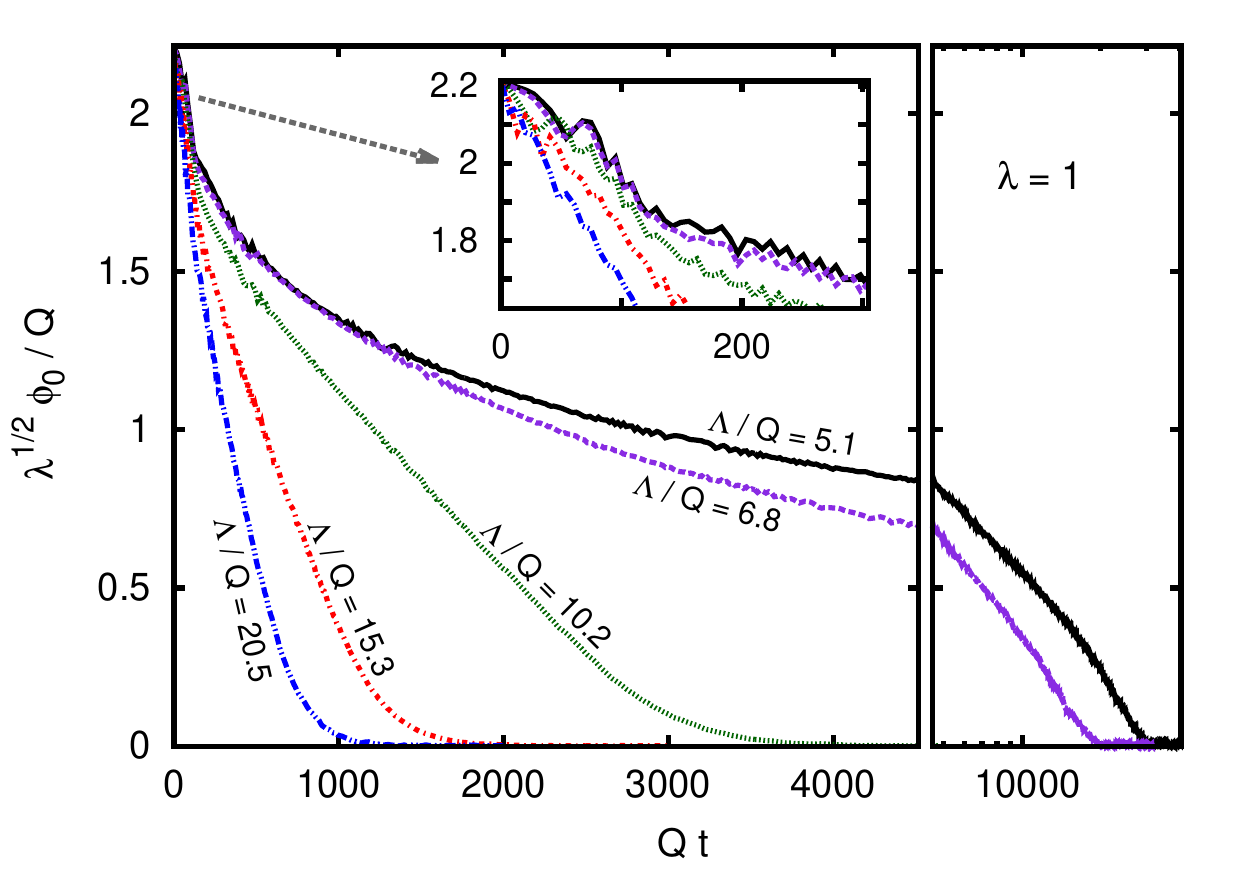}
 \caption{The rescaled field amplitude as a function of time for $\lambda=1$ and different lattice cutoffs. On the right hand side of the figure, a logarithmic time scale is used. The inset reveals a strong cutoff dependence of the results already at rather early times.}
\label{fig:condensatecomp}
\end{figure}

Since the range of vacuum modes included in this classical-statistical setup is controlled by the Rayleigh-Jeans cutoff, one expects the results to become strongly cutoff dependent once the unphysical decay of quantum modes sets in. We will now compare the dynamics for various values of the cutoff $\Lambda$, with the mass renormalization as outlined above. Fig.~\ref{fig:condensatecomp} shows the field amplitude as a function of time. One finds that the condensate becomes zero at a finite time while the decay time itself can be varied to practically any value in a vast range by changing $\Lambda$. Since the decay happens faster the larger the cutoff, the prediction for Bose condensation of Ref.~\cite{Epelbaum:2011pc} has to be considered as an artifact of the employed regularization.

\begin{figure}[t]
\centering
 \includegraphics[width=0.5\textwidth]{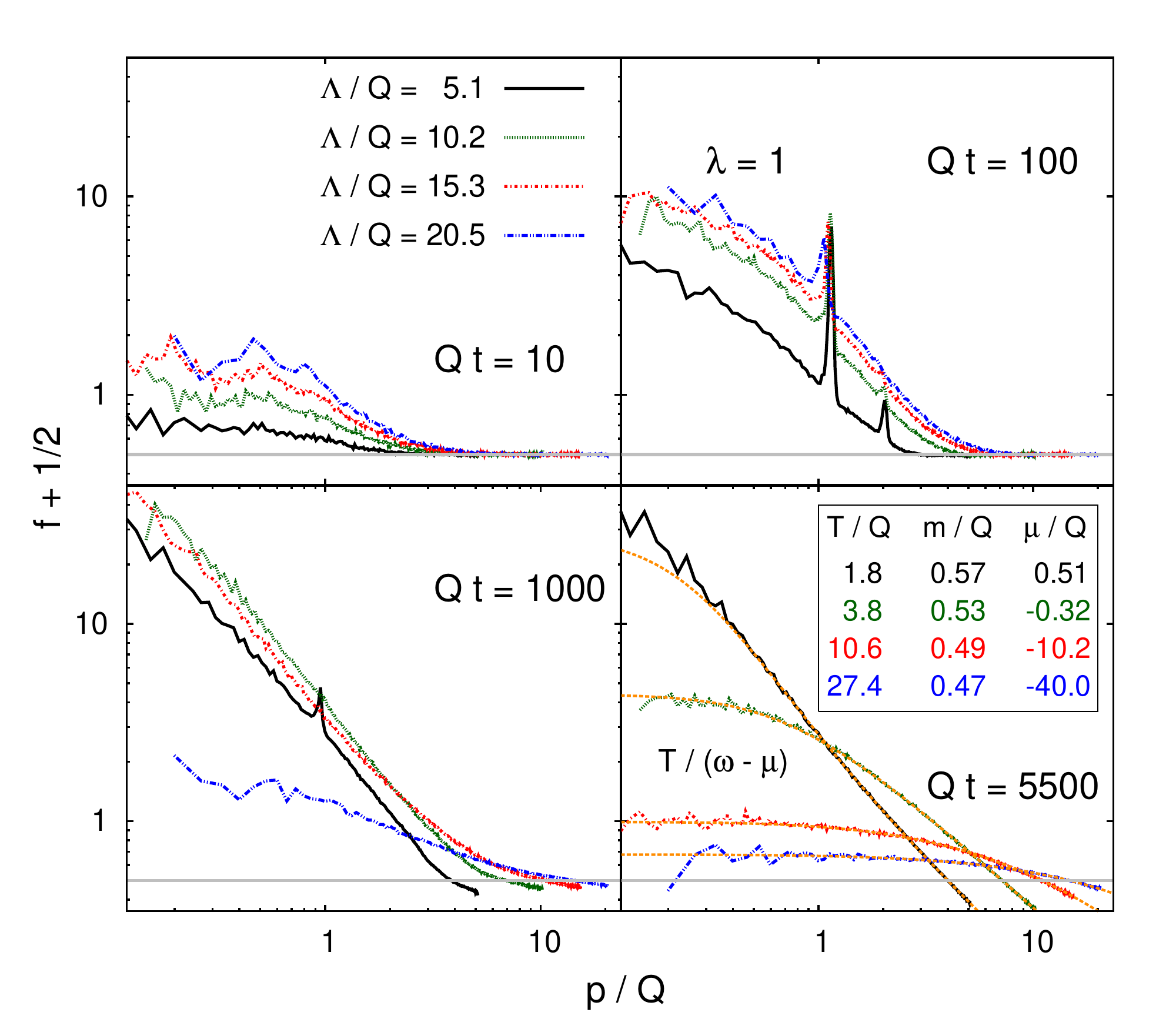}
 \caption{The occupation number distribution as a function of momentum for $\lambda = 1$ and different lattice cutoffs. Results are given for four different times. The table in the inset shows the dependence of the classical thermal fit parameters on the Rayleigh-Jeans cutoff.}
\label{fig:SpectraL1_2x2Plot}
\end{figure}

In Fig.~\ref{fig:SpectraL1_2x2Plot} we show the spectra $f(t,p) + 1/2$ for four different lattice cutoffs.  
Here a similar picture emerges. The spectra at four different times ranging from $Q t = 10$ to $Q t = 5500$ exhibit cutoff dependencies. The table with the extracted fit parameters at $Q t = 5500$ reveals that no sensible prediction independent of the cutoff can be made. 

\section{Conclusion}

Classical-statistical simulations provide an important tool for ab initio descriptions of nonequilibrium quantum dynamics. However, they have a well defined range of validity which is restricted to weak enough couplings. Since the coupling controls the relative size of the occupation number per mode as compared to the `quantum half' and the latter are cut off by the Rayleigh-Jeans regulator, the sizeable cutoff dependence indicates a breakdown of the classical-statistical approach for larger couplings. 
For the setup considered in the single component scalar field theory, we find that quantitative results can be obtained for $\lambda \lesssim 0.2$ for parametric times $t \ll Q^{-1} \lambda^{-5/4}$. In particular, our analysis shows that the results of Refs.~\cite{Epelbaum:2011pc,Dusling:2010rm} for $\lambda = 1$ are based on the application of the classical-statistical approximation beyond its range of validity. 

We have demonstrated that both the condensate driven as well as fluctuation dominated initial conditions belong to the basin of attraction of the same nonthermal fixed point within the range of validity of the classical-statistical description. This leads to the phenomenon of turbulent thermalization, where the evolution is characterized by universality. Our results show the existence of a dual cascade for the single component scalar field theory. While this has been intensively studied for $O(N)$ symmetric $N$-component field theories, with a detailed analytic understanding using large-$N$ techniques, our analysis points out that these provide also a remarkably accurate description of the universal properties for $N=1$.  

It would be extremely valuable to have a similar analysis for the range of validity of the classical-statistical approach describing the nonequilibrium quantum dynamics of longitudinally expanding systems. The latter are crucial for our understanding of ultrarelativistic collision experiments of heavy nuclei in the laboratory. Simulations in the theoretically clean weak coupling limit demonstrate the existence of a nonthermal fixed point in the space-time evolution of non-Abelian plasmas~\cite{Berges:2013eia,Berges:2013fga}. Alternative thermalization scenarios with the gauge coupling exceeding a certain strength~\cite{Gelis:2013rba} can be analyzed along the lines of the present work. \\

\acknowledgements

We thank Thomas Epelbaum and Fran\c cois Gelis for providing us with their lattice data to make detailed comparisons possible. We also thank Daniil Gelfand, Larry McLerran and D\'enes Sexty for helpful discussions and suggestions. This work was supported in part by the German Research Foundation (DFG). S.S and R.V. are supported by US Department of Energy under DOE Contract No.~DE-AC02-98CH10886. The numerical results presented in this work were obtained on the bwGRiD (\url{http://www.bw-grid.de}), member of the German D-Grid initiative, funded by the Ministry for Education and Research (BMBF) and the Ministry for Science, Research and Arts Baden-Wuerttemberg (MWK-BW).

\end{document}